\documentclass[aps,prb,twocolumn,superscriptaddress]{revtex4-1}
\usepackage{graphicx}
\usepackage{amssymb}
\usepackage{amsmath}
\usepackage{amsfonts}
\usepackage{color}

\begin{document}
\def\bfe{\mathbf{e}}
\def\bfx{\mathbf{x}}
\def\tr{\mathrm{tr}}
 
\title{Phase transition in the three-dimensional $O(N)\otimes O(M)$ model: a Monte Carlo study} 
\author{A.~O. Sorokin}
\email{aosorokin@gmail.com}
\affiliation{NRC ``Kurchatov Institute", Petersburg Nuclear Physics Institute, Gatchina 188300, Russia} 
\affiliation{St.Petersburg State University, 7/9 Universitetskaya nab., 199034 St.~Petersburg, Russia}

\begin{abstract}
Using Monte Carlo simulations, we consider the lattice version of the $O(N)\otimes O(M)$ sigma model for $2\leq M\leq4$ and $M\leq N \leq8$. We find a continuous transition for $N\geq M+4$. Estimates of the critical exponents for cases of second-order and weak first-order transitions are found. For $M=2$ our estimates of the exponents and marginal dimensionality $N_c^+(M)$ are in good agreement with the results of the non-perturbative renormalization group approach. For $M\geq2$ we find estimates of the exponents and marginal dimensionality between the values obtained in the first and second orders of the large-N expansion. To complete the picture, we also consider the usual $O(N)$ model ($M=1$).
\end{abstract}

\maketitle

\section{Introduction}

In the modern theory of critical phenomena, many analytical and numerical methods have been developed that allow to describe both qualitatively and quantitatively the critical behavior of various systems. Of course, each of these methods is associated with certain difficulties, both purely technical and lying in their justification and application area. As a rule, they allow to obtain acceptable quantitative estimates at least for systems from the universality class of the $O(N)$ model. From the point of view of  renormalization group (RG) approaches, the simplicity of the $O(N)$ model lies in the uniqueness of the coupling constant, i.e. there is only one non-trivial fixed point, which is IR attractive in $2<d<4$. For models with two or more coupling constants, the situation is more complicated, a stable fixed point may be absent, and then one observes a fluctuation-induced first-order transition. The properties of a RG-flow and possible fixed points are widely discussed in multiple coupling scalar theories such as general $N$-vector models \cite{Brezin74,Michel84,Osborn18,Rychkov19,Codello20}. The situation remains controversial even for one of the simplest generalizations of the $O(N)$ model, namely for the $O(N)\otimes O(M)$ model with two coupling constants.

The $O(N)\otimes O(M)$ model has arisen almost half a century ago in the context of studying transitions in spin systems with non-collinear ordering (such as helimagnets)\cite{Bak76,Garel76,Brazovskii76} and superfluid helium-3\cite{Jones76}. (See \cite{Delamotte04} for a review.) To date, this model has been considered in the framework of various approaches: $4-\varepsilon$ expansion \cite{Kawamura88, Kawamura90, Sokolov95, Pelissetto01, Calabrese04, Kompaniets20}, $1/N$ expansion \cite{Pelissetto01, Gracey02, Gracey02-2}, perturbative RG \cite{Sokolov94, Loison00, Pelissetto01-2, Pelissetto01-3, Sokolov02,Calabrese03,Parruccini03,Pelissetto04,Delamotte10}, pseudo-$\varepsilon$ expansion \cite{Calabrese04,Holovatch04}, $2+\epsilon$ expansion \cite{Azaria90,Azaria93,David96,Pelissetto01} (see also\cite{Hikami81} for the $N=M$ and $N=M+1$ cases), non-perturbative (functional) RG \cite{Zumbach93,Zumbach94,Zumbach94-2, Delamotte00,Delamotte03,Delamotte04,Delamotte16} (NPRG), and the conformal bootstrap (CB) program \cite{Nakayama14,Nakayama15,Henriksson20}.

Unexpectedly, the simplest and most considered case $M=2$ turns out to be the most controversial. (The case $M=1$ is the usual $O(N)$ model.) Moreover, this controversy relates to the most physically significant cases $N=2$ and $N=3$. Apart from these cases, discrepancies in the predictions of different approaches are reduced to quantitative estimates of critical exponents and the value $N_c^+(M,d)$ such that for $M\leq N<N_c^+(M,d)$ a stable fixed point is absent but appears for $N>N_c^+(M,d)$. Estimates of $N_c^+(M,3)$ obtained by various theoretical methods for $2\leq M\leq4$ are shown in table \ref{Table1}. However, the perturbative (fixed-dimension) RG computations performed at  six loops within the zero momentum massive scheme for $M=2$ predict the additional critical value $N_{c2}<N_c^+(2,3)$ below of which a stable fixed point reappears and exists for $N=2,\,3$ and 4 \cite{Sokolov02,Calabrese03,Pelissetto04}. (In this approach, $N_c^+(2,3)\approx6.4$ and $N_{c2}\approx5.7$.) Numerical analysis of the RG-flow geometry, based on the resummation of the 6-loop approximation for $\beta$-functions, suggests that this new fixed point is of the focus-type with a complex-valued correction-to-scaling exponents $\omega$. In contrast to the perturbative RG, the $4-\varepsilon$, pseudo-$\varepsilon$ expansions as well as the non-perturbative RG do not predict the appearance of such a fixed point.

If the perturbative RG would be the only method giving results contrary to other methods then one can appeal to his unreliability. In fact, this method is less rigorously justified than the $4-\varepsilon$ and $1/N$ expansions, if only because of the absence of a formal small expansion parameter, and its  results are high sensitive with respect to the resummation parameters. Nevertheless, this approach gives acceptable quantitative estimates of the critical behavior of the $O(N)$ model, and series obtained from the more reliable $4-\varepsilon$ and $1/N$ expansions are also only asymptotic with rather poor convergence properties for physically interesting values of $\varepsilon$ and $N$. In addition, the conformal bootstrap program  \cite{Nakayama14,Nakayama15,Henriksson20} also predicts the existence of a non-trivial fixed point below $N_c^+(2,3)$ with critical exponents in good agreement with the fixed-dimension perturbative results.

The conformal bootstrap determines the exact bound to the scaling dimensions of operators. These exclusion bounds may have kinks which as expected correspond to the position of actual exponents of the critical point.The main advantage of this method is that it is not based on series expansions and does not have convergence problems, contrary to RG approaches. The disadvantages of the method include the fact that it postulates scale invariance which is absent upon a first-order transition, and although even mild kinks can be interpreted as the scaling dimensions of the conformal field theory corresponding to the critical point, only the presence of kinks cannot serve as evidence that a transition is continuous.  In addition, the conformal bootstrap predicts an ordinary fixed point instead of the focus-type, so the situation for the $O(2)\otimes O(2)$ and $O(3)\otimes O(2)$ symmetry classes remains unclear.

\begin{table}[t]
\caption{\label{Table1} Numerical estimates of $N_c^+(M,3)$ for $M=2,\,3,\,4$ obtained by different approaches within various orders of perturbation theory and resummation procedures. PB --- Pad\'e--Borel; PBL --- Pa\'e–-Borel-–Leroy; DSIS --- direct summation of the inverse series; CB --- conform--Borel; P --- Pad\'e; LPA --- local potential approximation; LPA' is LPA with a moment-dependent anomalous dimension.}
\begin{tabular}{llll}
\hline
\hline
Method & $M=2$ & $M=3$ & $M=4$ \\
\hline
$4-\varepsilon$, $\mathcal{O}(\varepsilon^3)$, PB \cite{Sokolov95} & $3.39$ &  &  \\
$4-\varepsilon$, $\mathcal{O}(\varepsilon^3)$ \cite{Pelissetto01} & $5.3(2)$ & $9.1(9)$ & $12.1(1)$  \\
$4-\varepsilon$, $\mathcal{O}(\varepsilon^4)$, PB \cite{Kompaniets20} & $4.6(2.1)$ & $7.7(1.4)$ & $10.3(1.6)$ \\
$4-\varepsilon$, $\mathcal{O}(\varepsilon^5)$, PBL \cite{Calabrese04} & $5.47(7)$ & $\sim9$ &   \\
$4-\varepsilon$, $\mathcal{O}(\varepsilon^5)$, DSIS \cite{Calabrese04} & $6.1(2)$ & $9.6(4)$ & $12.7(7)$  \\
$4-\varepsilon$, $\mathcal{O}(\varepsilon^5)$, PB \cite{Kompaniets20} & $5.3(7)$ & $8.4(1.1)$ & $11.2(1.3)$  \\
$4-\varepsilon$, $\mathcal{O}(\varepsilon^6)$, PB \cite{Kompaniets20} & $5.8(8)$ & $9.3(5)$ & $12.3(6)$  \\
$4-\varepsilon$, $\mathcal{O}(\varepsilon^6)$, CB \cite{Kompaniets20} & $6.0(6)$ & $9.3(4)$ & $12.4(3)$ \\
$4-\varepsilon$, $\mathcal{O}(\varepsilon^6)$, DSIS & $5.9(2)$ & $9.2(4)$ & $12.2(5)$  \\
PRG, $\mathcal{O}(g^4)$, PB \cite{Sokolov94} & $3.91(1)$ &  &    \\
PRG, $\mathcal{O}(g^7)$, PB \cite{Calabrese03} & $6.4(4)$ & $11.1(6)$ & $14.7(8)$   \\
Pseudo-$\varepsilon$, $\mathcal{O}(\tau^6)$, P \cite{Holovatch04} & $6.23(21)$ &  &   \\
Pseudo-$\varepsilon$, $\mathcal{O}(\tau^6)$, P \cite{Calabrese04} & $6.22(12)$ & $9.9(3)$ & $13.2(6)$  \\
$1/N$, $\mathcal{O}(1/N)$ & $3.8$ & $5.2$ & $6.6$  \\
$1/N$, $\mathcal{O}(1/N^2)$ \cite{Pelissetto01} & $5.3$ & $7.3$ & $9.2$  \\
NPRG, LPA \cite{Zumbach93} & $4.7$ &  &   \\
NPRG, LPA' \cite{Delamotte16} & $5.24(2)$ &  &   \\
CB \cite{Nakayama14} &  & $\sim7$ &    \\
This work & $5.5(4)$ & $6.5(4)$ & $7.5(4)$ \\
\hline
\hline
\end{tabular}
\end{table}

Monte Carlo (MC) simulations of lattice models from these symmetry classes do not bring complete clarity to the problem. The main difficulty here is typical for models with several coupling constants, like the $O(N)\otimes O(M)$ model: lattice systems can undergo a first-order phase transition even if a stable fixed point exists on the RG diagram, but initial values of coupling constants locate outside the attraction region of this point.

Many different models have been considered using various MC algorithms (see \cite{Loison04,Delamotte04} for a review). The most famous of them is an antiferromagnet on a stacked-triangular lattice (STA). In early works for $N=2$ and $N=3$, both first-order and continuous phase transitions have been observed depending on models. Moreover, some models with second-order behavior demonstrate the universality. At that, the tendency towards continuous as well as universal behavior for the case $N=3$ is more pronounced. In fact, these finite-size lattice results can be explained in terms RG even if a stable fixed point is absent, but if a corresponding RG trajectory passes through a region characterized by a very slow evolution of RG parameters. Such a region may arise, e.g., if a fixed point has complex-valued coordinates with a small imaginary part.  Moreover, if this region is small and attracts trajectories starting from a wide set of initial values of RG parameters then the almost universal behavior is observed. It is this picture (including the tendency mentioned above) that is observed using the non-perturbative RG approach \cite{Zumbach93,Zumbach94,Zumbach94-2, Delamotte00, Delamotte03}.

Further numerical studies have confirmed the first-order transition scenario for the $N=2$ case (including STA\cite{Itakura03,Peles04,Diep08-1}, helimagnets\cite{Sorokin14}, the lattice version of the $O(2)\otimes O(2)$ model\cite{Itakura03,Okubo10}, the lattice version of the $O(2)\otimes O(2)$ sigma model\cite{Kunz93,Loison98}) as well as for the $N=3$ case (including STA\cite{Diep08}, helimagnets\cite{Sorokin14}, the lattice version of the $O(3)\otimes O(2)$ model\cite{Okubo10}, the lattice version of the $O(3)\otimes O(2)$ sigma model\cite{Loison99,Itakura03}). However, the recent study \cite{Kawamura19} of $N=3$ STA considering huge lattices finds a continuous transition. The authors \cite{Kawamura19} do not observe any double-peak structure the energy distribution in contrast to the results of the work \cite{Diep08}, and obtain an indication  the focus-type fixed point, namely, they obtain the complex-valued correction-to-scaling exponent. The authors note that the RG flow around the focus-like fixed point may temporarily moves from the potential stability region, that seems in finite-lattice studies as a first-order transition, sings of which disappear in the thermodynamic limit (so-called pseudo-first order). Apparently, further research is required to explain the inconsistencies in the results of works ref. \cite{Kawamura19} and ref. \cite{Diep08} using different MC algorithms.

Note that the perturbative RG predicts the focus-type fixed point for $N=3$ as well as for $N=2$, but for the later case MC simulations observe a first-order transition. In addition, the $N=2$ case of the $O(N)\otimes O(2)$ model has the same order parameter space $G/H=\mathbb{Z}\otimes SO(2)$ as the $N=2$ case of the Ising-$O(N)$ model (with three coupling constants), where a first-order transition is found \cite{Sorokin18,Sorokin19-2,Sorokin19-3}. Also note that the correction-to-scaling exponent can be complex-valued for a complex-valued fixed point, that can be observed for the pseudo-scaling behavior upon a weak first-order transition.

We have one more argument in favor of the scenario with a first-order transition at least for the $N=3$ case. There are topological excitations of the special type, namely so-called $\mathbb{Z}_2$-vortices, in the spectrum of the $O(3)\otimes O(2)$ model. We know that in two dimensions $\mathbb{Z}_2$-vortices can crucial change the critical behavior, in particular in the $O(3)\otimes O(3)$ model one observes a finite-temperature first-order transition instead of a Ising-like continuous one \cite{Sorokin17,Sorokin19-1}. In $2+\epsilon$ dimensions where a transition occurs at low temperature, $\mathbb{Z}_2$-vortices are associated in topologically neutral configurations, so a transition is of the second order from the universality class of the $O(4)$ model \cite{Azaria90,Azaria93}. One expects that the critical behavior changes at some finite $\epsilon<1$. So, we cannot exclude that at $\epsilon=1$ a transition becomes of the first order. 

Although the presence of topological defects of any types does not guarantee changes in the critical behavior, one can note that they are absent in the $O(N)\otimes O(2)$ model for $N\geq6$ that is close to the value $N_c^+(2,3)$. This coincidence is not reproduced for $M>2$ at least in RG approaches. However, the consistency of different RG methods in estimating the value $N_c^+(M,3)$ also deteriorates with increasing $M$. So, one should use a method without series expansions.

In this work we consider the $O(N)\otimes O(M)$ model, namely the lattice version of the $O(N)\otimes O(M)$ sigma model for $M=2,\,3,\,4$ and $N=M,\ldots,8$ using Monte Carlo simulations. Our results do not confirm the expectations of RG approaches that the first order of a transition becomes more pronounced with increasing $M$.

\section{Model and methods}

The $O(N)\otimes O(M)$ model is described by the Ginzburg -- Landau functional\cite{Kawamura90}
\begin{eqnarray}
    F=\int d^dx\left(\sum_i\Bigl((\partial_\mu\mathbf{\phi}_i)^2+r\mathbf{\phi}_i^2\Bigr)+\right.\nonumber\\
    \left.
    +u\Bigl(\sum_i\mathbf{\phi}_i^2\Bigr)^2+2v\sum_{i,j}\Bigl((\mathbf{\phi}_i \mathbf{\phi}_j)^2-\mathbf{\phi}_i^2\mathbf{\phi}_j^2\Bigr)\right),
    \label{GLW-model}
\end{eqnarray}
where $\phi_i$ is a $N$-component vector field, $i,\,j=1,\ldots,M$. The region of the potential stability with the non-collinear ground state in the broken symmetry phase $r<0$ is
\begin{equation}
    u>0,\quad v>0,\quad \frac{M}{M-1}u-v>0,
\end{equation} 
and the ground state is
\begin{equation}
\phi_i^2=\frac{-r}{2(Mu-(M-1)v)},\quad \phi_i\perp\phi_j.
\end{equation}
The order parameter $\Phi=(\phi_1,\ldots,\phi_M)$ is a $N\times M$-matrix. In the disordered phase, it is invariant under global $O(N)_L\otimes O(M)_R$ symmetry group acting correspondingly left and right on a matrix. In the ordered phase, the symmetry group is broken down to $O(N-M)_L\otimes O(M)_{\mathrm{diag}}$ subgroup. So, the order parameter space $G/H$ is a Stiefel manifold
\begin{equation}
\frac{O(N)_L\otimes O(M)_R}{O(N-M)_L\otimes O(M)_{\mathrm{diag}}}\approx \frac{O(N)}{O(N-M)}\equiv V_{N,M}.
\end{equation}
If we take the limits $u\to\infty$, $v\to\infty$ keeping $|\phi|=1$ and $u/u=\mathrm{const}$, we obtain the $O(N)\otimes O(M)$ sigma model. In this work, we consider this model on a lattice with the Hamiltonian
\begin{equation}
    H=-J\sum_{\bfx,\mu}\tr\,\Phi_\bfx^T\Phi_{\bfx+\bfe_\mu},\quad \mu=1,\ldots,3,
    \label{lattice-model}
\end{equation}
where $\bfe_\mu$ is a unit vector of a simple cubic lattice, $J>0$. Below, for brevity, we denote the $O(N)\otimes O(M)$ sigma model on a lattice as the $V_{N,M}$ model.

We investigate the $V_{N,M}$ model by Monte Carlo simulations using the Wollf cluster algorithm \cite{Wollf89}. We consider the cases $M=2,\,3,\,4$ and $N=M,\ldots,8$, thus we reproduce all known numerical results for the three-dimensional $V_{2,2}$\cite{Kunz93,Loison98}, $V_{3,2}$\cite{Diep94,Loison99,Itakura03}, $V_{3,3}$\cite{Diep94,Loison00-2}, $V_{4,3}$\cite{Loison00-2}, and $V_{4,4}$\cite{Loison00-2} models. In addition, we consider the simplest case $M=1$.

We use periodic boundary conditions and lattices with sizes $L=15,\,20,\,25,\,30,\,40,\,50,\,60,\,80$ for $M=1$, $L=15,\ldots,60$ for $M=2$ and $N>3$, and $L=15,\ldots,50$ for $M=3,\,4$ and $N>5$. In each simulation, $5\cdot10^5$ MC steps are made for thermalization, and $5\cdot10^6$ steps for calculation of averages.

A field configuration $\Phi_\bfx$ is defined by generalized Euler angles\cite{Hoffman72}. For the uniform distribution of a random direction on a hypersphere, it is necessary to define the following functions
\begin{equation}
f_n(\theta)=\int \sin^n\theta\,d\theta,\quad n=2,\ldots,6,
\end{equation}
and the inverse functions
\begin{equation}
\theta=f_n^{-1}(r),\quad r\in[0,1],\quad \theta\in[0,\pi],
\end{equation}
where $r$ is a random number. For the inverse functions, we use tables of values of size $6.4\cdot10^5$ and linear interpolation.

The order parameter is simply defined as
\begin{equation}
\mathbf{m}=\frac1{L^3}\sum_\bfx \phi_1(\bfx),\quad m=\sqrt{\mathbf{m}^2}.
\end{equation}
The estimation of the transition temperature is performed using the Binder cumulant crossing method \cite{Binder81}
\begin{equation}
    U=1-\frac{\langle m^4 \rangle}{3\langle m^2 \rangle^2}.
\end{equation}
Critical exponent $\nu$ is estimated using the following cumulants \cite{Ferrenberg91}:
\begin{equation}
V_n=\frac{\partial}{\partial(1/T)}\ln
\langle m^n\rangle=L^3\left(\frac{\left<m^n E\right>}{\left<m^n\right>}-\langle E\rangle\right),
\label{Vn}
\end{equation}
\begin{equation}
    \max\left(V_n\right)\sim L^{\frac1\nu}.
\end{equation}
Other exponents are estimated as follows:
\begin{equation}
    \left.m \right|_{T=T_c}\sim L^{-\frac\beta\nu},\quad
    \left.\chi\right|_{T=T_c}\sim L^{\frac\gamma\nu},
\end{equation}
where $\chi$ is the susceptibility
\begin{equation}
    \chi=\frac{L^3}{T}\left<m^2\right>,\quad T\geq T_c.
\end{equation}
Since we independently determine the exponents $\nu$, $\beta/\nu$, and $\gamma/\nu$ from our simulations, we can more accurately estimate the Fisher exponent $\eta$ using both scaling relations
\begin{equation}
	\eta=2-\frac\gamma\nu=2\frac{\beta}\nu-1.
\end{equation}   
It is very useful for a case of a weak first-order transition, where we do not observe a double-peak structure of the energy distribution. From the unitarity bound for the anomalous dimensions of the field $\Phi$, we have
\begin{equation}
\eta\geq0,\quad \frac{\beta}\nu\geq\frac12,\quad \frac\gamma\nu\leq2.
\end{equation}
Otherwise, we deals with a first-order transition.

\begin{table}[t]
\caption{\label{Table2}Homotopy groups\cite{Stiefel35,Whitehead45} of Stiefel manifolds $\pi_k(V_{N,M})$.}
\begin{tabular}{ccccc}
\hline
\hline
$G/H=V_{N,M}$ & $\pi_0(G/H)$ & $\pi_1(G/H)$ & $\pi_2(G/H)$ & $\pi_3(G/H)$\\
\hline
$V_{1,1}$ & $\mathbb{Z}_2$ & 0 & 0 & 0 \\
$V_{2,2}$ & $\mathbb{Z}_2$ & $\mathbb{Z}$ & 0 & 0 \\
$V_{3,3}$ & $\mathbb{Z}_2$ & $\mathbb{Z}_2$ & 0 & $\mathbb{Z}$ \\
$V_{4,4}$ & $\mathbb{Z}_2$ & $\mathbb{Z}_2$ & 0 & $\mathbb{Z}+\mathbb{Z}$ \\
$V_{N,N},\, N\geq5$ & $\mathbb{Z}_2$ & $\mathbb{Z}_2$ & 0 & $\mathbb{Z}$ \\
\hline
$V_{2,1}$ & 0 & $\mathbb{Z}$ & 0 & 0 \\
$V_{3,2}$ & 0 & $\mathbb{Z}_2$ & 0 & $\mathbb{Z}$ \\
$V_{4,3}$ & 0 & $\mathbb{Z}_2$ & 0 & $\mathbb{Z}+\mathbb{Z}$ \\
$V_{N,N-1},\, N\geq5$ & 0 & $\mathbb{Z}_2$ & 0 & $\mathbb{Z}$ \\
\hline
$V_{3,1}$ & 0 & 0 & $\mathbb{Z}$ & $\mathbb{Z}$ \\
$V_{4,2}$ & 0 & 0 & $\mathbb{Z}$ & $\mathbb{Z}+\mathbb{Z}$ \\
$V_{N,N-2},\,N\geq5$ & 0 & 0 & $\mathbb{Z}$ & $\mathbb{Z}$ \\
\hline
$V_{4,1}$ & 0 & 0 & 0 & $\mathbb{Z}$ \\
$V_{N,N-3},\,N\geq5$ & 0 & 0 & 0 & $\mathbb{Z}_2$ \\
\hline
$N\geq M+4\geq5$ & 0 & 0 & 0 & 0 \\
\hline
\hline
\end{tabular}
\end{table}
Since we are going to discuss the presence of topological excitations of any type, it is useful to know some topological properties of the order parameter space $G/H=V_{N,M}$. In general, a $q$-dimensional topological configuration is topologically protected in $d$-dimensions if the homotopy group $\pi_{d-q-1}(G/H)$ is non-trivial. So in three dimensions, a 2-dimensional configuration is a domain wall. Domain walls appear if the order parameter space has the form $G/H=G_d\otimes C$, where $G_d$ is a discrete group, and $C$ is a connected homogeneous space. 1-dimensional topological configurations are vortex tubes. Besides topological defects of these types, skyrmion-like configurations may be present if $\pi_d(G/H)$ is non-trivial. Necessary information about the topology of Stiefel manifolds is shown in Table \ref{Table2}. One can note that topological defects of any types are absent for $N\geq M+4$.

As we have discussed above, the presence of topological defects does not guarantee changes in the critical behavior, but we can formulate a criterion for when topological configurations make a significant contribution. For lattice models, one can define the (total) density of topological defects using the local definition of defects. This quantity contains two terms corresponding to free and associated defects: $\rho_\mathrm{total}=\rho_\mathrm{free}+\rho_\mathrm{pairs}$ for point-like defects, and $\rho_\mathrm{total}=\rho_\mathrm{infinite}+\rho_\mathrm{closed}$ for extended ones. In the ordered phase $\rho_\mathrm{free}$ and $\rho_\mathrm{infinite}$ tend to be zero, while in the disordered phase these quantities have some finite values, renormalized by critical fluctuations. Without fluctuations, the defect density has a jump at the transition point, and we deal with a first-order transition. In a case of strong fluctuations, the situation is more delicate. Point-like defects associated in pairs or closed extended defects have the topological charge of zero, so they are indistinguishable from ordinary non-topological excitations, but can screen the topological charge of free defects, making $\rho_\mathrm{free}$ or $\rho_\mathrm{infinite}$ finite in the ordered phase. So, the singularity of the topological defect density becomes softer or quite disappears. Since the internal energy is proportional to the total defect density\cite{Sorokin19}, a significant contribution of topological defects in the critical behavior means that the specific heat (as derivative of the internal energy with respect to temperature) has a singularity
\begin{equation}
C\sim(T-T_c)^{-\alpha},\quad \alpha>0.
\end{equation}
In particular, monopole-like configurations in the $O(3)$ model with $\alpha<0$ are not relevant to the critical behavior as discussed in refs.\cite{Holm94,Antunes02}.

\section{Results}

\subsection{$M=1$}

\begin{table}[t]
\caption{\label{TableTc}Critical temperature for the $V_{M,N}$ model.}
\begin{tabular}{lllll}
\hline
\hline
 & $M=1$ & $M=2$ & $M=3$ & $M=4$ \\
\hline
$N=1$ & $4.51150(4)$ &  &  &  \\
$N=2$ & $2.20163(5)$ & $2.444(1)$ & & \\
$N=3$ & $1.44295(5)$ & $1.53119(6)$ & $1.670(1)$  & \\
$N=4$ & $1.06855(5)$ & $1.11768(5)$ & $1.174(1)$  & $1.206(1)$ \\
$N=5$ & $0.84640(5)$ & $0.87840(7)$ & $0.91245(7)$  & $0.91936(8)$ \\
$N=6$ & $0.69998(5)$ & $0.72254(7)$ & $0.74602(8)$  & $0.75010(8)$ \\
$N=7$ & $0.59621(5)$ & $0.61286(8)$ & $0.63105(8)$  & $0.63296(8)$ \\
$N=8$ & $0.51902(5)$ & $0.53232(8)$ & $0.54481(8)$  & $0.54501(8)$ \\
\hline
\hline
\end{tabular}
\end{table}
\begin{table}[b]
\caption{\label{TableM1}Critical exponents for the case $M=1$.}
\begin{tabular}{llll}
\hline
\hline
$N$ & $\nu$ & $\beta$ & $\gamma$\\
\hline
$1$ & $0.630(5)$ & $0.327(2)$ & $1.236(10)$ \\
$2$ & $0.672(5)$ & $0.348(3)$ & $1.320(10)$ \\
$3$ & $0.712(6)$ & $0.370(4)$ & $1.396(12)$ \\
$4$ & $0.750(7)$ & $0.388(4)$ & $1.474(14)$ \\
$5$ & $0.760(7)$ & $0.392(4)$ & $1.496(14)$ \\
$6$ & $0.784(7)$ & $0.406(4)$ & $1.541(14)$ \\
$7$ & $0.830(8)$ & $0.433(5)$ & $1.624(16)$ \\
$8$ & $0.850(8)$ & $0.436(5)$ & $1.678(16)$ \\
\hline
\hline
\end{tabular}
\end{table}
We consider the case $M=1$ for two reasons. First, it allows us to test our modeling technique, that is especially important for large $N$. For most values of $N$, the critical temperatures and exponents are known more accurately then in this work. We just fill in some gaps. Second, we use the case $M=1$ to fit of the critical temperature as a function of $N$ and $M$.

Our results on the estimation of the critical temperature are shown in Table \ref{TableTc}, and the critical exponents in the case $M=1$ are shown in Table \ref{TableM1}.

The simplest fitting of the inverse critical temperature is
\begin{equation}
\frac{J}{T_c}\equiv K_c\approx 0.2440835N-0.0335268.
\end{equation}
A more general fit using the results for $M>1$ is as follows:
\begin{equation}
K_c\approx K_1 N+K_0,
\end{equation}
where
\begin{eqnarray}
K_1=0.247208-0.004056M+0.001239M^2,\nonumber\\
K_0=0.020598-0.055813M+0.001772M^2.\nonumber
\end{eqnarray}

\subsection{$M=2$}

\begin{table*}
\caption{\label{TableM2}Critical exponents for the case $M=2$.}
\begin{tabular}{clllllll}
\hline
\hline
\,\,$V_{N,M}$\,\, &  & $\nu$ & $\alpha$ & $\beta$ & $\gamma$ & $\beta/\nu$ & $\eta$\\
\hline
& This work & $0.572(6)$ & $0.284(18)$ & $0.276(4)$ & $1.165(14)$ & $0.482$ & $-0.037(14)$ \\
$V_{4,2}$ & $1/N$, $\mathcal{O}(1/N)$ \cite{Pelissetto01} & $0.676$ & $-0.03$ & $0.365$ & $1.297$ & $0.541$ & $0.081$ \\
\hline
& This work & $0.621(7)$ & $0.137(21)$ & $0.308(8)$ & $1.246(17)$ & $0.496$ & $-0.008(15)$ \\
$V_{5,2}$ & PRG, $\mathcal{O}(g^4)$, PB \cite{Loison00} & $0.565$ & $0.305$ & $0.300$ & $1.095$ & $0.531$ & $0.063$ \\
& $1/N$, $\mathcal{O}(1/N)$ \cite{Pelissetto01} & $0.676$ & $-0.03$ & $0.365$ & $1.297$ & $0.541$ & $0.081$ \\
\hline
& This work & $0.686(7)$ & $-0.058(21)$ & $0.354(8)$ & $1.35(2)$ & $0.516$ & $0.032(17)$  \\
& MC, STA \cite{Loison00} & $0.700(11)$ & $–0.100(33)$ & $0.359(14)$ & $1.383(36)$ & $0.505$ & $0.025(20)$  \\
& $4-\varepsilon$, $\mathcal{O}(\varepsilon^6)$, CB \cite{Kompaniets20} & $0.65(2)$ & $0.05$ & $0.34$ & $1.27(3)$ & $0.523$ & $0.047(3)$  \\
$V_{6,2}$ &PRG, $\mathcal{O}(g^4)$, PB \cite{Loison00} & $0.575$ & $0.275$ & $0.302$ & $1.121$ & $0.525$ & $0.051$  \\
& $1/N$, $\mathcal{O}(1/N)$ \cite{Pelissetto01} & $0.730$ & $-0.19$ & $0.390$ & $1.410$ & $0.534$ & $0.068$  \\
& $1/N$, $\mathcal{O}(1/N^2)$ \cite{Pelissetto01} & $0.633$ & $0.10$ & $0.336$ & $1.227$ & $0.531$ & $0.061$  \\
& NPRG, LPA' \cite{Delamotte16} & $0.695(5)$ & $-0.09$ & $0.362$ & $1.36$ & $0.521$ & $0.042(2)$  \\
\hline
& This work & $0.739(7)$ & $-0.217(21)$ & $0.381(8)$ & $1.456(20)$ & $0.515$ & $0.030(17)$  \\
& $4-\varepsilon$, $\mathcal{O}(\varepsilon^5)$, PBL \cite{Calabrese04} & $0.71(4)$ & $-0.13$ & $0.37$ & $1.39(6)$ & $0.52$ & $0.042(3)$  \\
& $4-\varepsilon$, $\mathcal{O}(\varepsilon^6)$, CB \cite{Kompaniets20} & $0.713(8)$ & $-0.139$ & $0.373$ & $1.396(14)$ & $0.523$ & $0.045(3)$  \\
& PRG, $\mathcal{O}(g^4)$, PB \cite{Loison00} & $0.566$ & $0.303$ & $0.295$ & $1.108$ & $0.521$ & $0.042$  \\
$V_{7,2}$ & PRG, $\mathcal{O}(g^7)$, CM \cite{Calabrese03} & $0.68(2)$ & $-0.04$ & $0.354$ & $1.31(5)$ & $0.521$ & $0.042(2)$  \\
& $1/N$, $\mathcal{O}(1/N)$ \cite{Pelissetto01} & $0.768$ & $-0.305$ & $0.406$ & $1.492$ & $0.523$ & $0.058$  \\
& $1/N$, $\mathcal{O}(1/N^2)$ \cite{Pelissetto01} & $0.697$ & $-0.09$ & $0.367$ & $1.357$ & $0.523$ & $0.053$  \\
& NPRG, LPA' \cite{Delamotte16} & $0.735(5)$ & $-0.21$ & $0.382$ & $1.44$ & $0.520$ & $0.039(2)$  \\
\hline
& This work & $0.771(8)$ & $-0.313(24)$ & $0.400(8)$ & $1.516(20)$ & $0.518$ & $0.034(20)$  \\
& $4-\varepsilon$, $\mathcal{O}(\varepsilon^5)$, PBL \cite{Calabrese04} & $0.75(4)$ & $-0.25$ & $0.40$ & $1.45(6)$ & $0.53$ & $0.067(3)$  \\
& $4-\varepsilon$, $\mathcal{O}(\varepsilon^6)$, CB \cite{Kompaniets20} & $0.745(11)$ & $-0.235$ & $0.388$ & $1.461(17)$ & $0.521$ & $0.042(2)$  \\
$V_{8,2}$ & PRG, $\mathcal{O}(g^4)$, PB \cite{Loison00} & $0.616$ & $0.152$ & $0.319$ & $1.211$ & $0.518$ & $0.035$  \\
& PRG, $\mathcal{O}(g^7)$, CM \cite{Calabrese03} & $0.71(1)$ & $-0.13$ & $0.369$ & $1.40(2)$ & $0.520$ & $0.039(1)$  \\
& $1/N$, $\mathcal{O}(1/N)$ \cite{Pelissetto01} & $0.797$ & $-0.39$ & $0.419$ & $1.554$ & $0.525$ & $0.051$  \\
& $1/N$, $\mathcal{O}(1/N^2)$ \cite{Pelissetto01} & $0.743$ & $-0.23$ & $0.389$ & $1.451$ & $0.524$ & $0.047$  \\
\hline
\hline
\end{tabular}
\end{table*}
\begin{table*}
\caption{\label{TableM3}Critical exponents for the case $M=3$.}
\begin{tabular}{clllllll}
\hline
\hline
\,\,$V_{N,M}$\,\, &  & $\nu$ & $\alpha$ & $\beta$ & $\gamma$ & $\beta/\nu$ & $\eta$\\
\hline
& This work & $0.564(18)$ & $0.31(6)$ & $0.264(12)$ & $1.164(40)$ & $0.468$ & $-0.063(40)$ \\
$V_{6,3}$ & $1/N$, $\mathcal{O}(1/N)$ \cite{Pelissetto01} & $0.640$ & $0.08$ & $0.349$ & $1.222$ & $0.545$ & $0.09$ \\
\hline
& This work & $0.635(8)$ & $0.095(24)$ & $0.328(9)$ & $1.249(24)$ & $0.516$ & $0.033(20)$ \\
$V_{7,3}$ & $1/N$, $\mathcal{O}(1/N)$ \cite{Pelissetto01} & $0.691$ & $-0.073$ & $0.372$ & $1.329$ & $0.539$ & $0.077$ \\
\hline
& This work & $0.701(14)$ & $-0.10(5)$ & $0.373(14)$ & $1.358(40)$ & $0.531$ & $0.063(40)$ \\
$V_{8,3}$ & $1/N$, $\mathcal{O}(1/N)$ \cite{Pelissetto01} & $0.730$ & $-0.19$ & $0.390$ & $1.410$ & $0.534$ & $0.068$ \\
& $1/N$, $\mathcal{O}(1/N^2)$ \cite{Pelissetto01} & $0.641$ & $0.076$ & $0.341$ & $1.242$ & $0.532$ & $0.064$ \\
\hline
\hline
\end{tabular}
\end{table*}
\begin{table*}
\caption{\label{TableM4}Critical exponents for the case $M=4$.}
\begin{tabular}{clllllll}
\hline
\hline
\,\,$V_{N,M}$\,\, & & $\nu$ & $\alpha$ & $\beta$ & $\gamma$ & $\beta/\nu$ & $\eta$\\
\hline
$V_{6,4}$ & This work & $0.54(4)$ & $0.38(12)$ & $0.27(4)$ & $1.08(11)$ & $0.50$ & $0.00(10)$ \\
\hline
& This work & $0.612(17)$ & $0.16(5)$ & $0.306(12)$ & $1.225(40)$ & $0.49$ & $-0.002(20)$ \\
$V_{7,4}$ & $1/N$, $\mathcal{O}(1/N)$ \cite{Pelissetto01} & $0.614$ & $0.16$ & $0.337$ & $1.169$ & $0.548$ & $0.10$ \\
\hline
& This work & $0.643(12)$ & $0.07(4)$ & $0.347(10)$ & $1.236(30)$ & $0.54$ & $0.078(30)$ \\
$V_{8,4}$ &$1/N$, $\mathcal{O}(1/N)$ \cite{Pelissetto01} & $0.662$ & $0.013$ & $0.359$ & $1.27$ & $0.542$ & $0.084$ \\
\hline
\hline
\end{tabular}
\end{table*}
\begin{figure}[t]
    \center
    \includegraphics[scale=0.30]{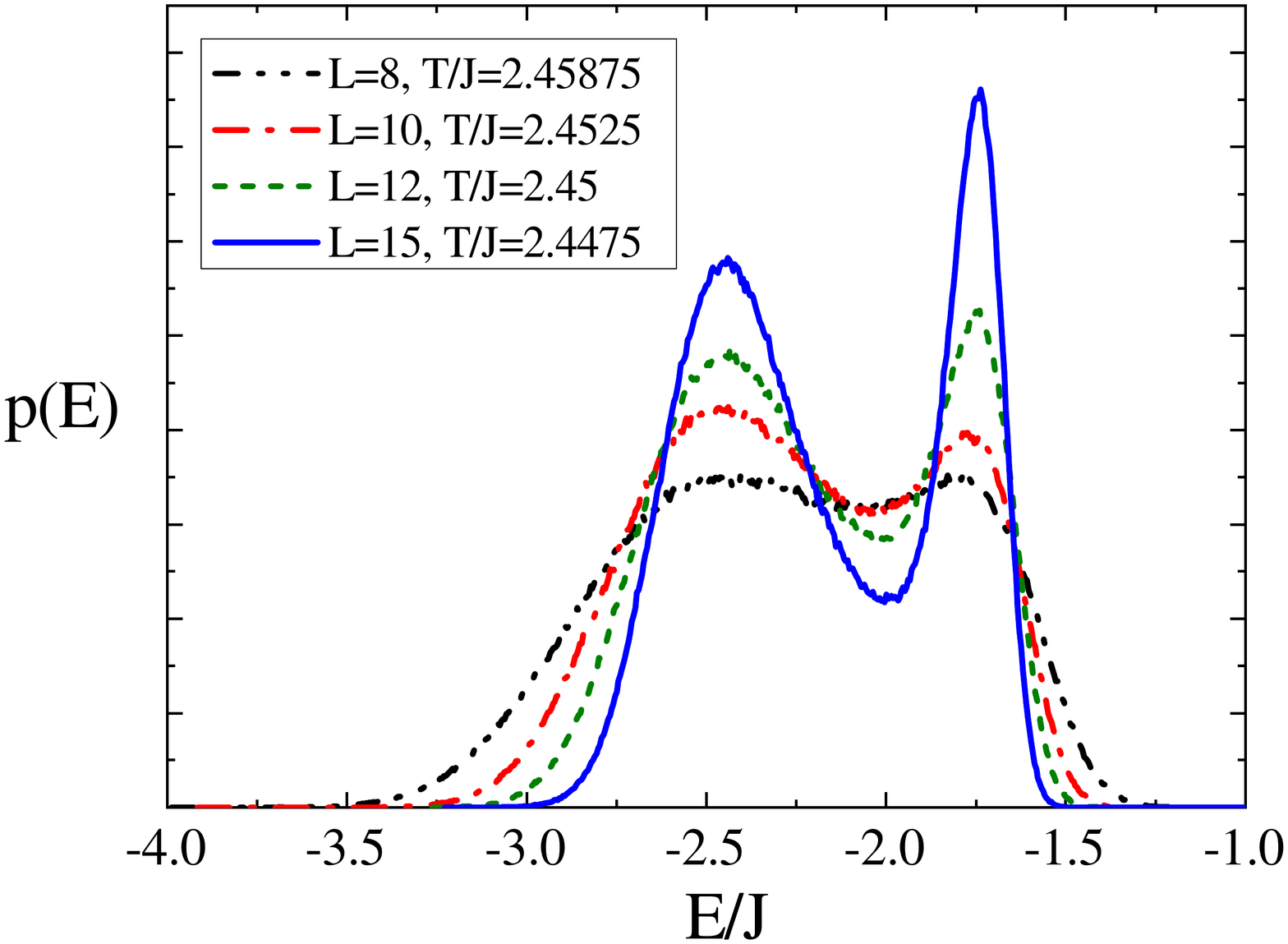}%
    \caption{\label{fig1} Internal energy distribution in the $V_{2,2}$ model}
\end{figure}%
\begin{figure}[t]
    \center
    \includegraphics[scale=0.30]{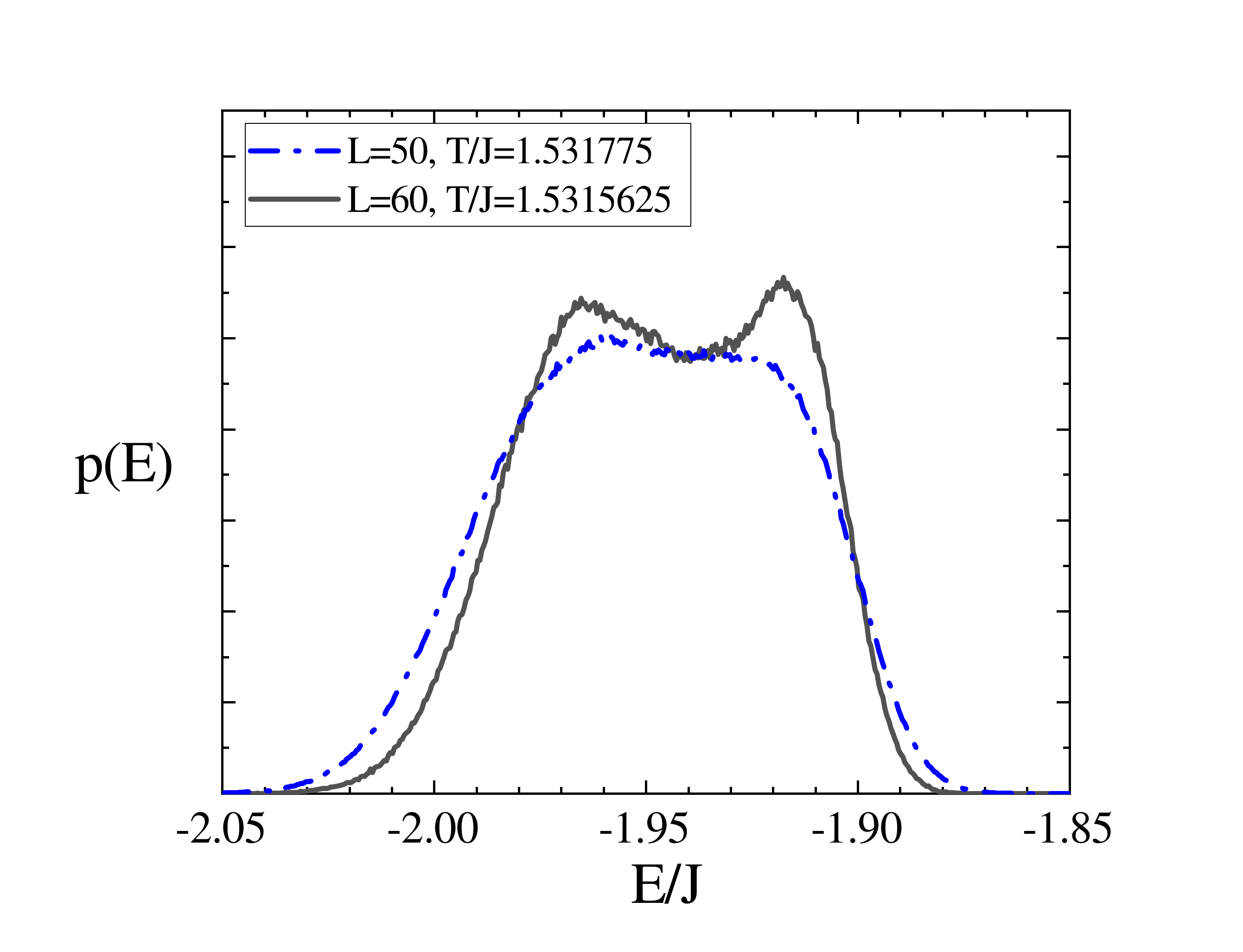}%
    \caption{\label{fig2} Internal energy distribution in the $V_{3,2}$ model}
\end{figure}%
As expected\cite{Loison98,Loison99,Itakura03}, we find a transition of the pronounced first order for the $V_{2,2}$ and $V_{3,2}$ models. Figs. \ref{fig1} and \ref{fig2} show a typical double-peak structure of the internal energy distributions. For the $V_{4,2}$ and $V_{5,2}$ models, we do not observe such a structure up to $L=60$. So, the pseudo-scaling exponents can be estimated, and we find that the Fisher exponent is negative $\eta<0$ (see Table \ref{TableM2}). We interpret this as a weak first-order transition.

For the $V_{6,2}$, $V_{7,2}$ and $V_{8,2}$ models, we find a second-order phase transition. It should be especially noted that our results are in good agreement with the results for $N=6$ stacked-triangular antiferromagnet\cite{Loison00}, as well as with the study within the framework of the non-perturbative RG approach\cite{Delamotte16} for $N=6$ and $N=7$.

The simplest fitting of the inverse critical temperature for $M=2$ is
\begin{equation}
K_c\approx 0.244812 N-0.082677.
\end{equation}

\subsection{$M=3$}

\begin{figure}[t]
    \center
    \includegraphics[scale=0.30]{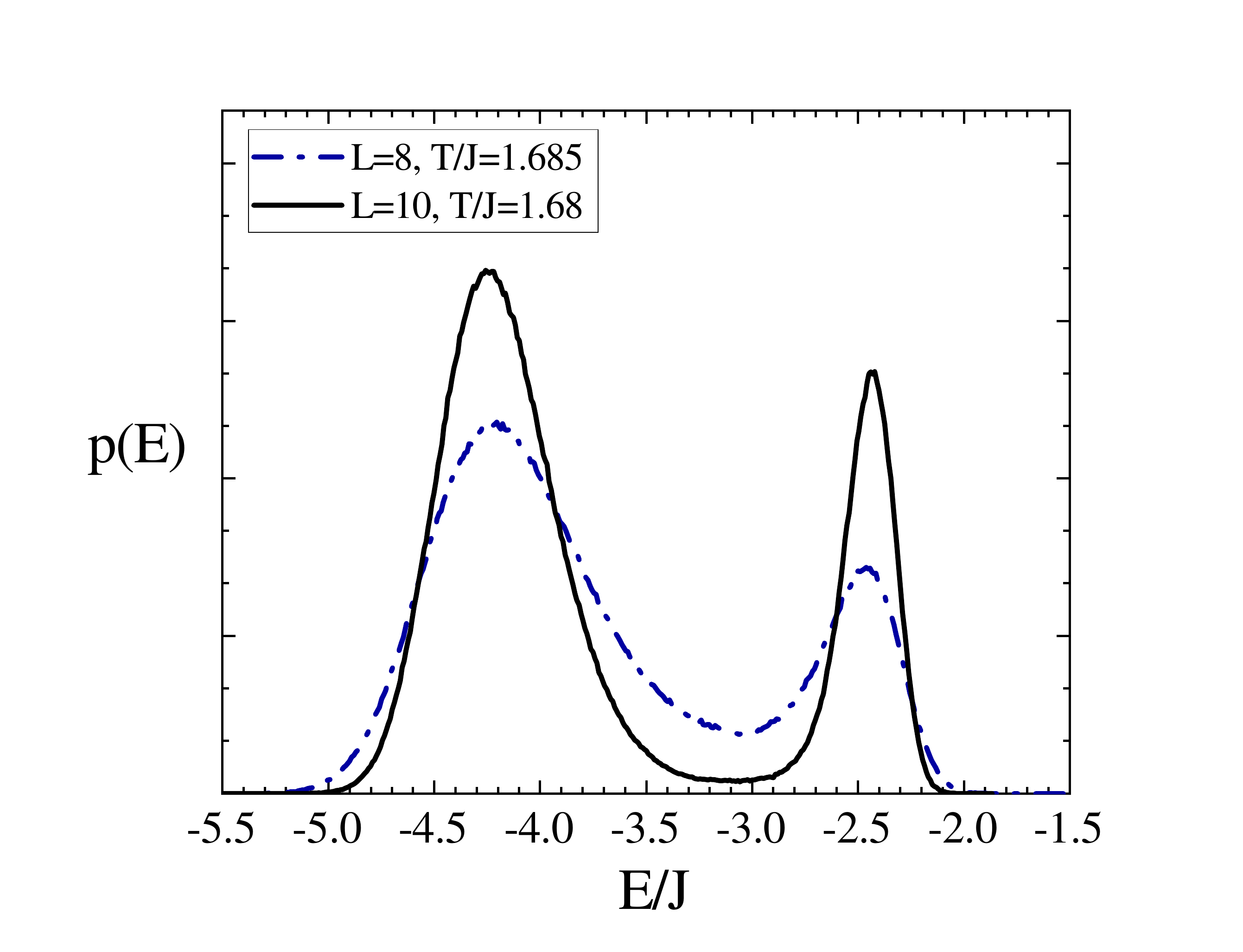}%
    \caption{\label{fig3} Internal energy distribution in the $V_{3,3}$ model}
\end{figure}%
\begin{figure}[t]
    \center
    \includegraphics[scale=0.30]{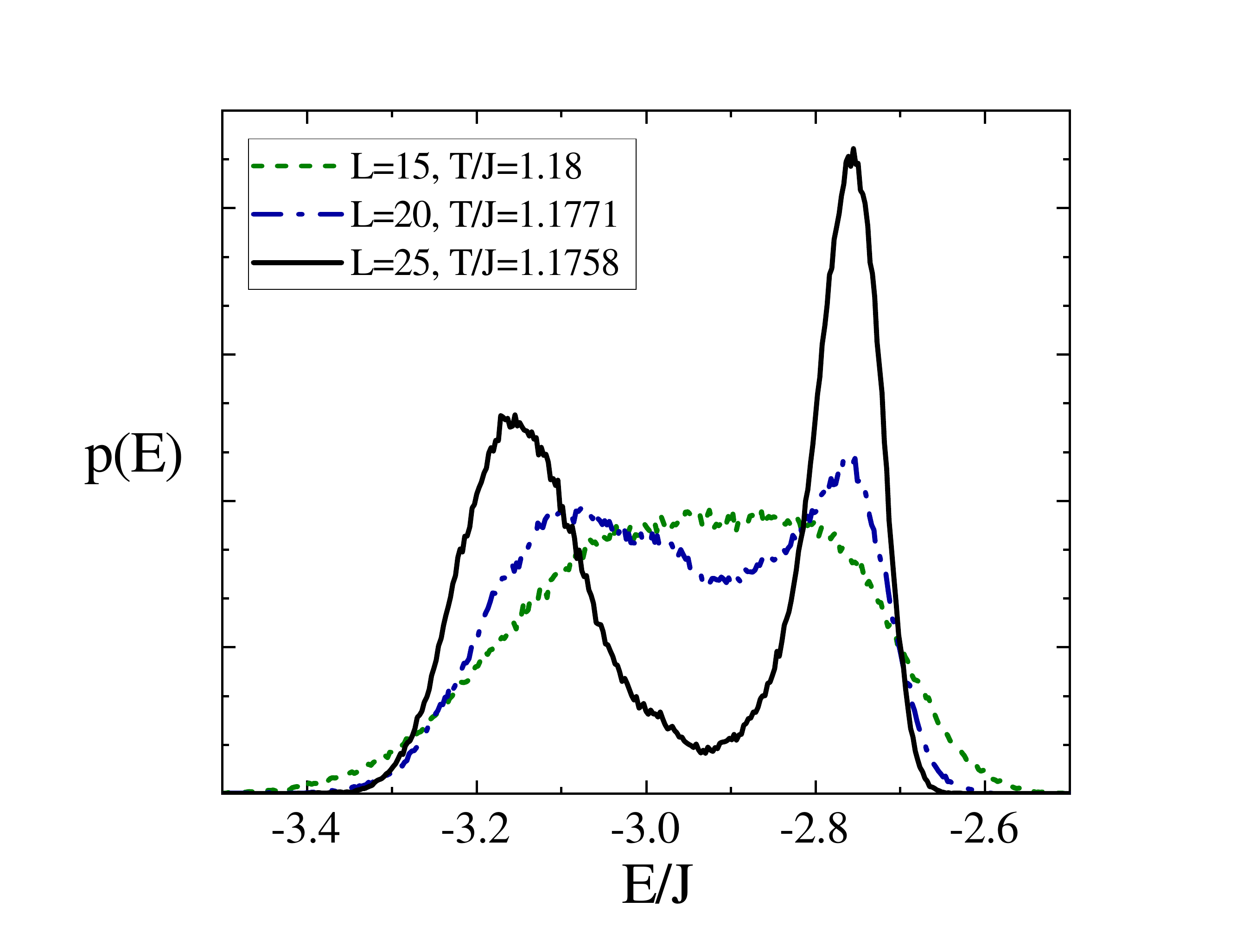}%
    \caption{\label{fig4} Internal energy distribution in the $V_{4,3}$ model}
\end{figure}%
\begin{figure}[t]
    \center
    \includegraphics[scale=0.30]{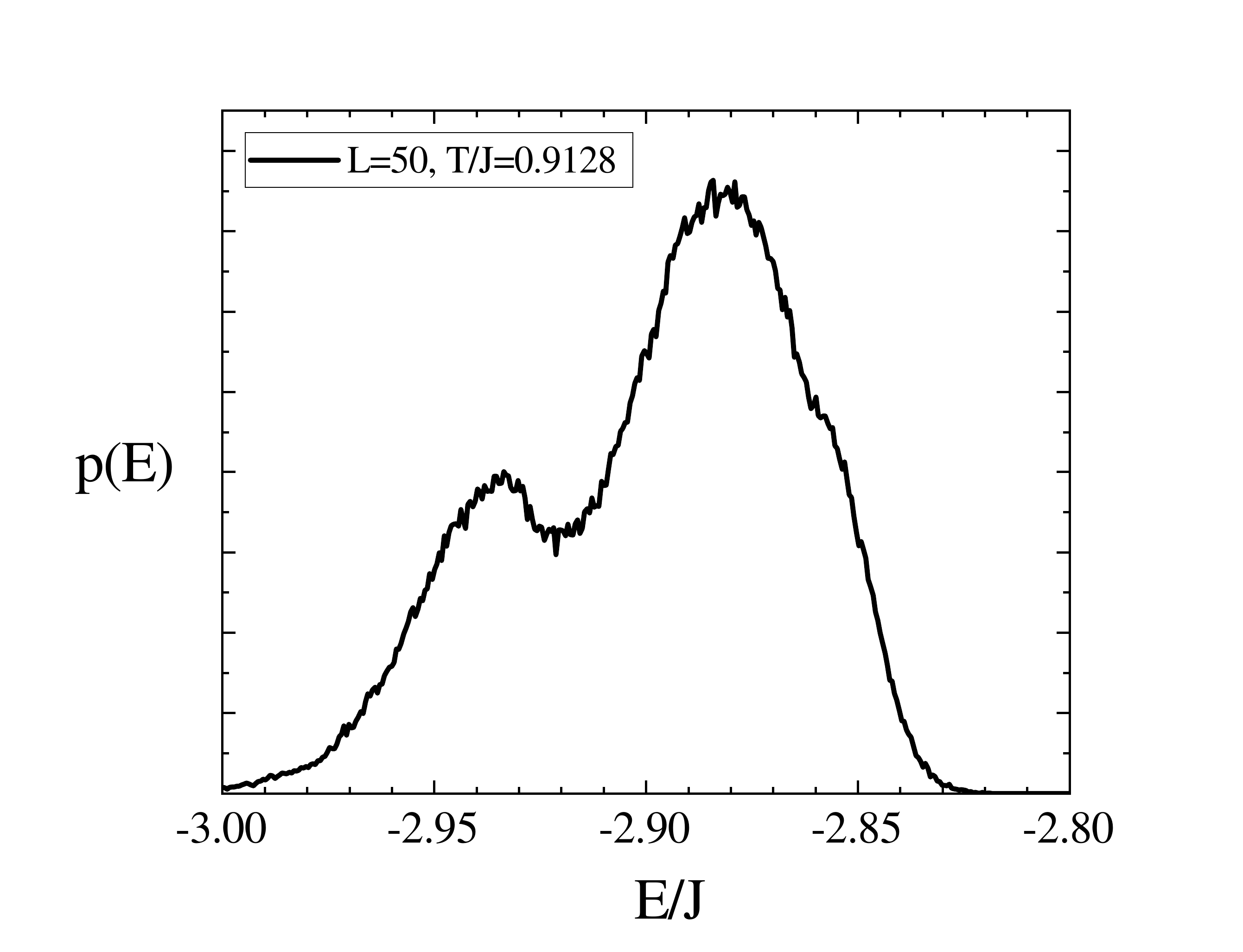}%
    \caption{\label{fig5} Internal energy distribution in the $V_{5,3}$ model}
\end{figure}%
We reproduce the results of ref.\cite{Loison00-2} and find a distinct first-order transition for the $V_{3,3}$ and $V_{4,3}$ models (see figs. \ref{fig3} and \ref{fig4}). However, for the case $M=3$, we obtain the same result for the $V_{5,3}$ model (fig. \ref{fig5}). In the case of the $V_{6,3}$ model, we find a weak first-order transition with the negative value of $\eta$ (see Table \ref{TableM3}).

Somewhat more unexpectedly, we observe a second-order transition for the $V_{7,3}$ and $V_{8,3}$. This contradicts the results of the perturbative RG as well as the $4-\varepsilon$ and pseudo-$\varepsilon$ expansions.

Again, the simplest fitting of the inverse critical temperature is
\begin{equation}
K_c\approx 0.245994N-0.135670
\end{equation}

\subsection{$M=4$}
					
\begin{figure}[t]
    \center
    \includegraphics[scale=0.30]{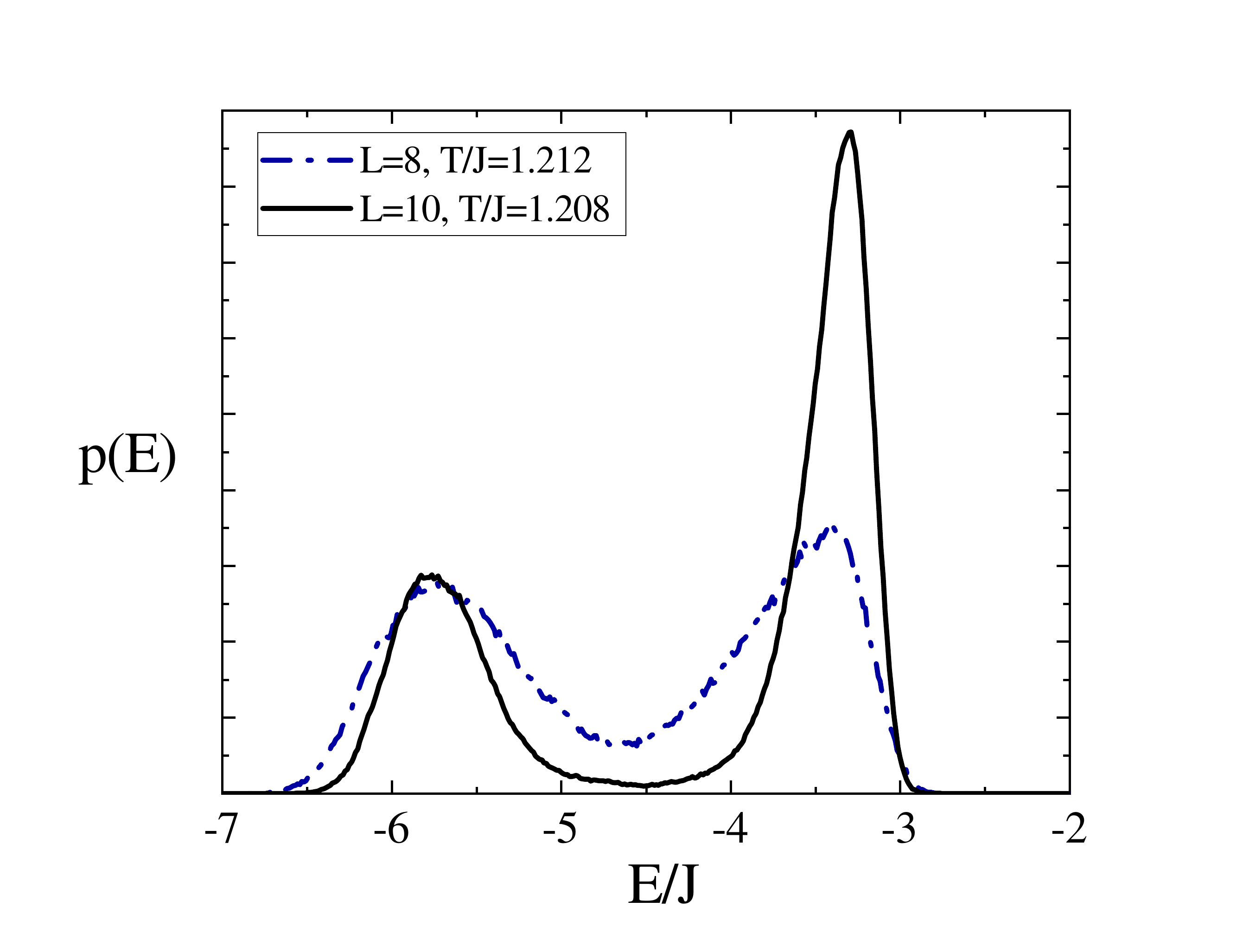}%
    \caption{\label{fig6} Internal energy distribution in the $V_{4,4}$ model}
\end{figure}%
\begin{figure}[t]
    \center
    \includegraphics[scale=0.30]{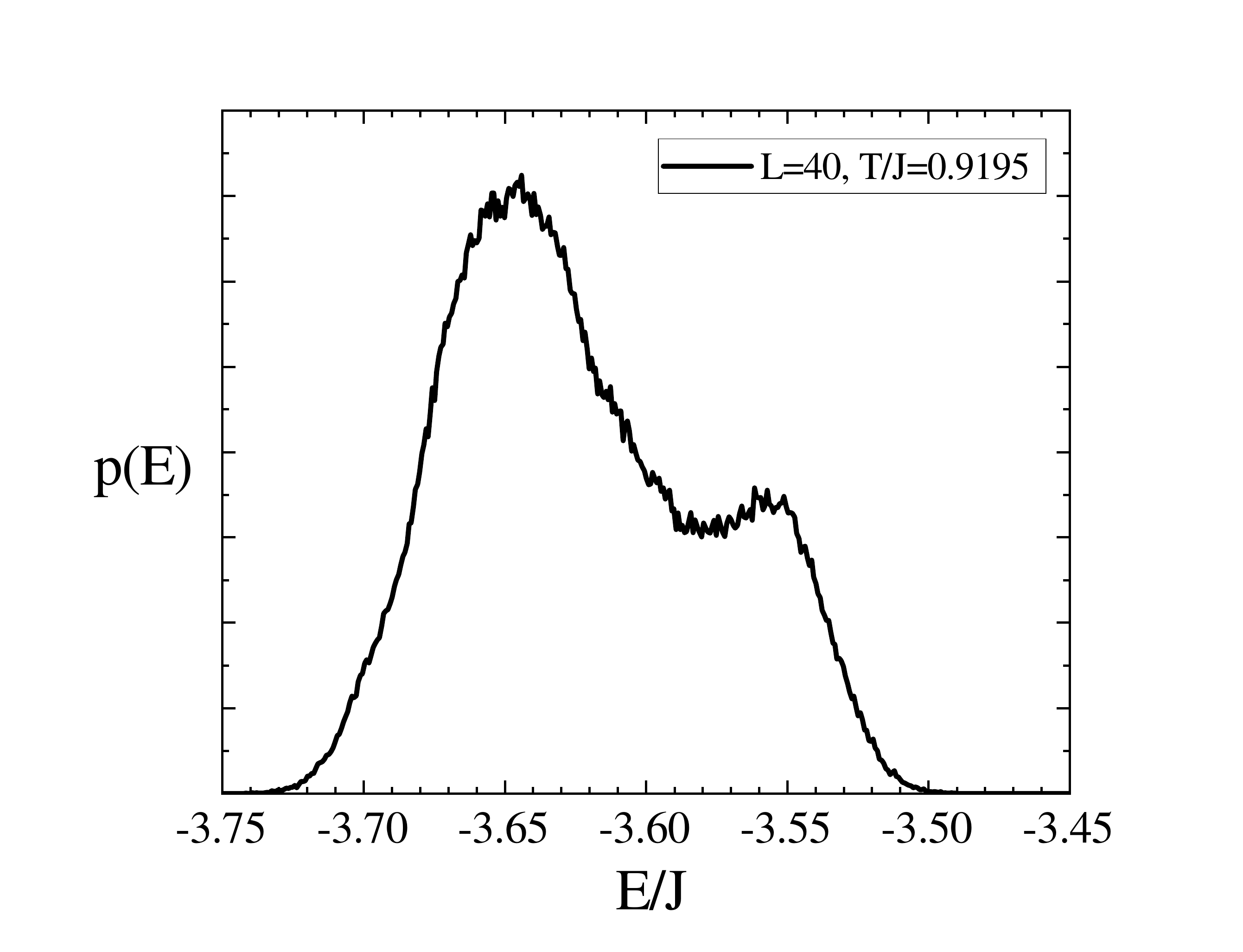}%
    \caption{\label{fig7} Internal energy distribution in the $V_{5,4}$ model}
\end{figure}%
In this case, we also reproduce the results of ref.\cite{Loison00-2} and find a distinct first-order transition for the $V_{4,4}$ model (see fig. \ref{fig6}). A distinct first-order transition occurs also in the $V_{5,4}$ model (fig. \ref{fig7}). However, for the $V_{6,4}$ and $V_{7,4}$ models, we find the weak first order (see Table \ref{TableM4}).

The $V_{8,4}$ model has a continuous transition.

The simplest fitting of the inverse critical temperature is
\begin{equation}
K_c\approx 0.250345N-0.169116 
\end{equation}

\section{Conclusion}

\begin{table}[t]
\caption{\label{TableOrder}Order of a transition in the $V_{N,M}$ model. Weak I order means that we do not observe a double-peak structure of the energy distribution, but $\eta<0$. The lattice size $L$ indicates that we do not observe a double-peak structure on smaller lattices.}
\begin{tabular}{lccc}
\hline
\hline
 & $M=2$ & $M=3$ & $M=4$\\
\hline
$N=2$ & I, $L\geq8$ & & \\
$N=3$ & \,\, I, $L\geq50$\,\, & I, $L\geq8$ & \\
$N=4$ & weak I  & I,$L\geq20$ & I, $L\geq8$ \\
$N=5$ & weak I  & \,\, I, $L\geq50$\,\, & \,\, I, $L\geq40$\,\, \\
$N=6$ & II  & weak I & weak I \\
$N=7$ & II  & II & weak I \\
$N=8$ & II  &  II &  II \\
\hline
\hline
\end{tabular}
\end{table}    
We performed extensive numerical investigation of the $V_{N,M}$ model, and obtained a few rather interesting results. We found the value of $N_c^+(M,3)$ is less than predicted by the perturbative RG and the $4-\varepsilon$ expansion for $M>2$. Although it may be a coincidence, but we found that a transition is of the second order for cases where topological defects are absent. The results of determining the order of a transition are collected in the Table \ref{TableOrder}. It would be interesting to compare values of the critical exponents for $M>2$ with predictions of the non-perturbative RG and the conformal bootstrap program.

Also we find that for $M\geq2$ the estimates of the exponents and marginal dimensionality $N_c^+(M,3)$ lie between the values obtained in the first and second orders of the large-N expansion without resummation. Possibly, the resummation of the large-N series improves the agreement with the results of the numerical analysis, but for this it is useful to calculate the third-order corrections, that is quite a difficult task.

\begin{acknowledgments}
This work was supported by the Theoretical Physics and Mathematics Advancement Foundation 'BASIS' (project No. 19-1-3-38-1). 
\end{acknowledgments}

\end{document}